\documentclass[preprint,prd,noshowpacs]{revtex4}
\usepackage{amssymb}
\usepackage{amsmath}
\usepackage{amsfonts}

\begin{document}

\title{Hawking Radiation as a Mechanism for Inflation}

\author{Sujoy Kumar Modak}
\email{sujoy@bose.res.in}
\affiliation{S.~N.~Bose National Centre for Basic Sciences, JD Block, Sector III, Salt Lake, Kolkata-700098, India}
\author{Douglas Singleton}
\email{dougs@csufresno.edu}
\affiliation{Physics Department, CSU Fresno, Fresno, CA 93740-8031 USA}

\date{\today}

\begin{abstract}
The Friedman-Robertson-Walker (FRW) space-time exhibits particle creation similar to Hawking radiation of a black hole. 
In this essay we show that this FRW Hawking radiation leads to an effective 
negative pressure fluid which can drive an inflationary 
period of exponential expansion in the early Universe. Since the Hawking temperature of the FRW space-time decreases as the
Universe expands this mechanism naturally turns off and the inflationary stage transitions to a power law expansion associated 
with an ordinary radiation dominated Universe.     
~\\
~\\
~\\
~\\
~\\
~\\
~\\
~\\
~\\
~\\
{\it Awarded ``Honorable Mention" for the Gravity Research Foundation 2012 Awards for Essays on Gravitation}
\end{abstract}

\maketitle

In this essay we propose a mechanism for inflation which has 
a short, early exponential expansion phase, followed by a transition to power law expansion with an exponent
of $\frac{1}{2}$ as expected for a Universe dominated by ordinary radiation. The mechanism is based on the
Hawking radiation associated with a Friedman-Robertson-Walker (FRW) space-time. The idea that particle creation
(Hawking radiation in this case) can influence cosmological evolution has a long history \cite{schrodinger}, \cite{parker}. 
The Hawking radiation of a space-time is closely connected with the thermodynamics of the space-time so we begin 
with the first law of thermodynamics,
\begin{eqnarray}
dQ=d(\rho V)+pdV ~. \label{flaw}
\end{eqnarray}
The left hand side denotes the heat ($Q$) change during the cosmic time $t$ to $t+dt$, $\rho$ is the 
energy density, $V$ is the volume and $p$ is the thermodynamic pressure. Dividing (\ref{flaw}) by $dt$, 
gives the relation between rate of changes in heat, energy and volume of the universe,
\begin{eqnarray}
\frac{dQ}{dt}=\frac{d}{dt}(\rho V)+p \frac{dV}{dt}. \label{dflaw}
\end{eqnarray}
In standard cosmological models the Universe is taken to be a {\it closed, adiabatic} system so that $dQ=0$. 
In this case the second law of thermodynamics, $dQ=TdS$, ensures that during this infinitesimal 
cosmic time interval entropy does not change.  However, if one allows particle creation from the gravitational field, 
in an irreversible manner, the universe is an open, adiabatic system. Thus heat transfer is allowed and entropy 
production is possible. This idea was first proposed in \cite{prigo}.

Here we consider a specific mechanism of particle creation:  the {\it Hawking effect} \cite{hawking} as 
applied to the early stages of a FRW space-time. The goal is to obtain inflation
without having to take recourse to an inflaton field with a tuned potential. 

The homogeneous, isotropic FRW line element is
\begin{eqnarray}
ds^2=-c^2dt^2+a^2(t)\left[\frac{dr^2}{1-kr^2}+r^2(d\theta^2+\sin ^2\theta d\phi^2) \right],
\label{frw}
\end{eqnarray}
where $a(t)$ is the scale factor and $k=0, \pm 1$ tells one if the spatial curvature of the Universe is flat ($k=0$), open
($k=-1$) or closed ($k=+1$). For this metric the Einstein field equations yield
the following two, non-trivial equations for the $\mu=\nu=0$ and $\mu=\nu=1,2,3 $ components respectively
\begin{equation}
3\frac{\dot a^2}{a^2} +3 \frac{k c^2}{a^2} = \frac{8\pi G \rho}{c^2} ~~ , ~~~~
2\frac{\ddot{a}}{a}+\frac{\dot a^2}{a^2}+\frac{kc^2}{a^2}=-\frac{8\pi G}{c^2}p ~, \label{00}
\end{equation}
where $\rho$ is energy density and $p$ pressure. 

Since the FRW universe is time-dependent, the definition of the cosmological event horizon is subtle. 
One can define the apparent horizon knowing the local 
properties of the space-time. In order to do this one can rewrite (\ref{frw}) in the following form \cite{sp-kim}
\begin{eqnarray}
ds^2=h_{ab}dx^adx^b+\tilde r^2(d\theta^2 +\sin^2\theta d\phi^2)
\end{eqnarray}
where, $x^a=(t,r)$ and $h_{ab}=$ diag$(-c^2,a^2/(1-kr^2))$ and $\tilde r= a (t) r$. 
The position of the apparent horizon is given by the root (${\tilde r}_A$) of the equation
$h^{ab}\partial_a \tilde r\partial_b \tilde r=0$. This is found to be
\begin{equation}
{\tilde r}_A =\frac{c}{\sqrt{H^2+\frac{kc^2}{a^2}}} \label{rah}
\end{equation}
where the Hubble parameter is defined as $H=\frac{\dot a}{a}$.

Using the above one can find the Hawking temperature of the apparent horizon \cite{sp-kim} 
\begin{equation}
T= \frac{\hbar c \kappa}{2 \pi k_B} =  \frac{\hbar c}{2 \pi k_B} 
\left( \frac{1}{{\tilde r}_A} \right) \left| 1-\frac{{\dot {\tilde r}}_A}{2 H {\tilde r}_A} \right|
 ,\label{htemp1}
\end{equation}
where $\kappa =\frac{1}{{\tilde r}_A} \left| 1-\frac{{\dot {\tilde r}}_A}{2 H {\tilde r}_A} \right|$ is the surface gravity.
During an inflationary phase the Universe's scale factor takes the form $a(t) \propto \exp( {constant \times t})$ so that
$H=\frac{\dot a}{a} = constant$. If the $constant$ satisfies $H^2 \gg c^2/a^2$ (later we show this is the case
during inflation) we have from (\ref{rah}) ${\tilde r}_A \approx \frac{c}{H} = constant$ and ${\dot {\tilde r}}_A  \approx 0$. 
Thus the temperature in (\ref{htemp1}) simplifies \cite{frw-hawking}
\begin{equation}
T=\frac{\hbar\sqrt{H^2+kc^2/a^2}}{2\pi k_B} \approx \frac{\hbar H }{2\pi k_B}.\label{htemp}
\end{equation}
In the final approximation we are again assuming $H^2 \gg c^2/a^2$ which we justify later.

At this point we note that the direction of radiation flux of the Hawking radiation for the apparent horizon in 
an FRW space-time is the opposite to that of a black hole event horizon. For black holes, the created particles 
escape outside the event horizon towards asymptotic infinity, while for the apparent horizon of FRW space-time 
the created particles come inward from the horizon. 
Due to the isotropy of FRW space-time, the radiation is isotropic from all directions. 
The net result is an  effective power gain in the Universe, given by the Stephan-Boltzman (S-B) radiation 
law {\footnote{Briefly the distinction between the radiation from black holes and from FRW space-times is as follows: 
for black holes $P$ is negative and they lose power during Hawking evaporation while for 
the FRW space-time $P$ is positive and so the Universe gains energy.}}
\begin{equation}
P=+\frac{dQ}{dt}=\sigma A_H T^4, \label{power}
\end{equation}
where $\sigma=\frac{\pi^2 k_B^4}{60\hbar^3 c^2}$ is the S-B constant and $A_H$ is the area of the
apparent horizon. Now substituting (\ref{power}) into (\ref{dflaw}) and using (\ref{htemp}) we obtain
\begin{equation}
\frac{d}{dt}(\rho V)+p \frac{dV}{dt}=\sigma A_H \left(\frac{\hbar H}{2\pi k_B}\right)^{4}.
\label{genrl}
\end{equation}
With the curvature term ignored (i.e. $k=0$) the volume and area 
of the Universe are $V=\frac{4\pi}{3} {\tilde r} _A ^3$ and $A_H=4\pi {\tilde{r}}_A ^2$ 
respectively \cite{narli, paddy}. Using $V, A_H$ and ${\tilde{r}}_A \approx c/H$ in (\ref{genrl}), we obtain
\begin{equation}
\dot{\rho} +3 (\rho + p )\frac{\dot a}{a} = \frac{3\sigma}{c}\left(\frac{\hbar}{2\pi k_{B}}\right)^4 H^5. \label{neq}
\end{equation}
We now rewrite (\ref{neq}) using the first equation in (\ref{00}) and $H=\frac{\dot a}{a}$ as 
\begin{equation}
\frac{\dot{\rho}}{\rho}+3(1+\omega)\frac{\dot a}{a} = 3 \omega_c (t) \frac{\dot a}{a} ~, \label{neq1}
\end{equation} 
where the equation of state for ordinary matter is taken as $p = \omega \rho$ and the
time dependent equation of state parameter due to particle creation is denoted $\omega_c (t)$ and is given by 
\begin{equation}
\omega_c (t) = \alpha \rho (t) ~~~~~~ \text{and} ~~~~~~ \alpha = \frac{\hbar G^2}{45 c^7}= 4.8 \times 10^{-116} (J/m^3)^{-1}. \label{alpha}
\end{equation} 
Moving the $\omega _c (t)$ term in (\ref{neq1}) from the rhs to the lhs one can see that this particle creation term acts like
a negative pressure. For the present Universe this term is negligible. The present value of the energy density of the 
Universe is $\rho_{0}=8.91\times 10^{-10}~J/m^3$ so that the $\omega _c (t_0) = \alpha \rho _0$ term on the rhs of (\ref{neq1})
is effectively zero. Thus this effect can not explain the current 
accelerated expansion of the Universe -- one still needs dark energy. However in the early Universe $\rho$ 
can be large enough so that the particle creation pressure on rhs of (\ref{neq1}) dominates.
Remembering that $\omega _c (t) \propto \rho (t)$ we solve the first order differential equation 
(\ref{neq1}) to find the modified energy density
\begin{equation}
\rho=\frac{Da^{-3(1+\omega)}}{1+(\frac{\alpha D}{1+\omega})a^{-3(1+\omega)}} \rightarrow 
\frac{D a^{-4}}{1+\frac{3\alpha D}{4}a^{-4}} = \frac{D}{a^4+\frac{3\alpha D}{4}},\label{nend}
\end{equation}
where $D$ is a constant and in the last equations we have taken the equation of state of the ordinary matter
to be $\omega=\frac{1}{3}$ (i.e. the equation of state of ordinary radiation) since we want the early Hawking radiation
inflation phase to be followed by a Universe dominated by ordinary radiation. 
The dimensions of $D$ depend on the value of the equation of state parameter $\omega$.
Note, in the classical limit ($\hbar\rightarrow 0$), $\alpha \rightarrow 0$ (the FRW Hawking radiation
effect turns off) and (\ref{nend}) gives the well known form $\rho \propto a^{-3(1+\omega)}$ for
a Universe dominated by ordinary matter with an equation of state $p = \omega \rho$.

There are two limits of (\ref{nend}): (i) $\alpha D \gg a^4$ so that $\rho \approx 4/(3 \alpha)=constant$ and the 
Hawking radiation effect dominates; (ii) $a^4 \gg \alpha D$ so that $\rho \approx D/a^4$ and the
energy density is of a Universe dominated by ordinary radiation. In both cases (i) and (ii) one can say that the Universe
is radiation dominated, but for case (i) this means FRW Hawking radiation while for case (ii) this means
ordinary radiation.

For case (i) where $\rho \approx 4/(3 \alpha) = constant$ one can insert this into the first Friedman equation in  
(\ref{00}) and integrate to obtain
\begin{equation}
a(t) \propto e^{H t} ~~~~~~ \text{where} ~~~~~
H= \frac{\dot a}{a} = \left(\frac{32 \pi G}{9c^2\alpha}\right)^{1/2} \approx 10^{45} \text{sec}^{-1} ~. \label{inflation-era}
\end{equation}
For this value of $H$ the condition $H^2 \gg k \frac{c^2}{a^2}$, which was assumed after (\ref{genrl}),
is valid so long as $a$ is larger than the Planck size (e.g. for $a \approx 100 \times l_{pl} = 10^ {-33} m$ 
while from (\ref{inflation-era}) $H ^2 \approx 10^{90}$ and $\frac{c^2}{a^2} \approx 10 ^{83}$). Note that the assumption 
that ${\dot {\tilde r}}_A  \approx 0$ is  justified {\it a posterior} since from (\ref{rah}) 
${{\tilde r}}_A  \approx \frac{1}{H}$ and during the inflationary phase $H$ is approximately constant.

Next, we look at case (ii) where $\rho \approx D/a^4$. This is just the case of a Universe
dominated by ordinary radiation. Inserting this $\rho$ into the first equation in (\ref{00}) and integrating gives
\begin{equation}
a(t) \approx \left( \frac{32 \pi G D}{3 c^2} \right)^{1/4}  t^{1/2} ~, \label{pl}
\end{equation}
which is the usual $t^{1/2}$ power law expansion for a Universe dominated by ordinary radiation. 
Thus after the inflationary stage given by (\ref{inflation-era}) there is a transition to a Universe dominated
by ordinary radiation (\ref{pl}). 

In summary our proposed FRW Hawking radiation mechanism for inflation gives rise to an effective negative pressure
term in the energy density evolution equations (\ref{neq}) and (\ref{neq1}). This leads to a modification of the 
energy density as a function of time given in (\ref{nend}). Looking at limiting cases of (\ref{nend})
we find that in the early Universe one has inflationary, exponential expansion (\ref{inflation-era}) which transitions 
to $t^{1/2}$ power law expansion (\ref{pl}) indicative of a Universe dominated by ordinary radiation. 
In this picture inflation is due to a quantum effect rather than being due to a scalar field with a tuned potential.

Finally we speculate that this model for inflation may not only explain why inflationary expansion turns off,
but also why it turns on. In discussions of black hole evaporation it has been suggested that near the end
stages of evaporation, when the black hole approaches the Planck scale, that Hawking radiation may stop. One concrete
example of this is the scenario in non-commutative geometry \cite{nicolini} where the Hawking temperature and 
evaporation rate go to zero as the black hole shrinks past some small cut-off length. Similarly, for a 
FRW space-time one can speculate that the associated FRW Hawking radiation might be turned off
when it is smaller than some cut-off scale. Only when the size of the FRW Universe enlarges past this cut-off 
size will the FRW Hawking radiation, and the associated exponential expansion given in (\ref{inflation-era}), turn on. 

{\bf Acknowledgment:} SKM thanks the American Physical Society for providing a travel grant through 
Indo-US Physics Student Visitation Program and CSU Fresno for providing
research amenities where this work was initiated. He is also supported by the Senior Research Fellowship form CSIR, Govt. of India.

\end{document}